\documentclass[%
reprint,
superscriptaddress,
amsmath,amssymb,
aps,
]{revtex4-1}

\usepackage{graphicx}
\usepackage{dcolumn}
\usepackage{bm}
\usepackage{color}

\begin{document}
	
\preprint{APS/123-QED}
	
\title{Coexistence of pseudospin- and valley-Hall-like edge states in a photonic crystal with $C_{3v}$ symmetry}
	
\author{Menglin L. N. Chen}
\affiliation{%
Department of Electrical and Electronic Engineering, The University of Hong Kong, Hong Kong
}%
\author{Li Jun Jiang}
\email{jianglj@hku.hk}
\affiliation{%
Department of Electrical and Electronic Engineering, The University of Hong Kong, Hong Kong
}%
\author{Zhihao Lan}
\email{z.lan@ucl.ac.uk}	
\affiliation{%
Department of Electronic and Electrical Engineering, University College London, United Kingdom
}%
\author{Wei E. I. Sha}
\email{weisha@zju.edu.cn}	
\affiliation{%
Key Laboratory of Micro-nano Electronic Devices and Smart Systems of Zhejiang Province, College of Information Science and Electronic Engineering, Zhejiang University, Hangzhou 310027, China
}%

\date{\today}

\begin{abstract}
	
We demonstrate the coexistence of pseudospin- and valley-Hall-like edge states in a photonic crystal with $C_{3v}$ symmetry, which is composed of three interlacing triangular sublattices with the same lattice constants. By tuning the geometry of the sublattices, three complete photonic band gaps with nontrivial topology can be created, one of which is due to the band inversion associated with the pseudospin degree of freedom at the $\Gamma$ point and the other two due to the gapping out of Dirac cones associated with the valley degree of freedom at the $K, K'$ points. The system can support tri-band pseudospin- and valley-momentum locking edge states at properly designed domain-wall interfaces. Furthermore, to demonstrate the novel interplay of the two kinds of edge states in a single configuration, we design a four-channel system, where the unidirectional routing of electromagnetic waves against sharp bends between two routes can be selectively controlled by the pseudospin and valley degrees of freedom. Our work combines the pseudospin and valley degrees of freedom in a single configuration and may provide more flexibility in manipulating electromagnetic waves with promising potential for multiband and multifunctional applications.

\end{abstract}

\maketitle

\textit{Introduction.---} Over the past few years, topological photonics, inspired by the development of topological states of matter in condensed matter physics~\cite{topo_wen}, such as, quantum Hall states~\cite{QH_klitzing17}, topological insulators~\cite{kane2010colloquium, sczhang2011TI}, has become a rapidly emerging area that holds great promise for the realization of robust control of electromagnetic (EM) waves~\cite{review_RMP19}. Various phenomena originally discovered in condensed matter physics, such as quantum Hall~\cite{qh80}, quantum spin Hall~\cite{kane_mele1,kane_mele2} and quantum valley Hall~\cite{dixiao2007valley} effects, are mimicked using photonic crystals (PCs) since the behavior of photons in PCs follows similar pattern to that of electrons in quantum systems. Indeed, PCs have been found to be powerful artificial media due to the flexiblity in manipulating EM waves~\cite{joannopoulos1997PCs}. Following the early success in implementing quantum Hall states using PCs~\cite{haldane08, Soljacic08, Soljacic09}, photonic systems that emulate the quantum spin Hall effects exploiting the polarization degeneracy between transverse-electric (TE) and transverse-magnetic (TM) modes have been extensively studied~\cite{Khanikaev2013photonic, chan_nc14, Ma2015guiding, chen_pnas, Khanikaev2016robust, tetm_apl19, Sievenpiper_lpr19}. A different proposal utilizing only the TM modes but exploiting the underlying crystalline symmetry of conventional dielectric PCs has successfully constructed a pair of pseudospin-up and -down photonic states based on the two inequivalent two-dimensional (2D) irreducible representations of the $C_{6v}$ symmetry group~\cite{huxiao2015scheme}. This proposal has been verified experimentally both in microwave and optical regimes~\cite{qsh_18PRL, qsh_19PRL, qsh_20sciadv, qsh_20nanoletter}. Recently, the valley degree of freedom associated with energy band maxima and minima at the $K, K'$ points of the Brillouin zone (BZ) of hexagonal lattices~\cite{dixiao2007valley} has also been introduced to photonics~\cite{Ma2016all}, which gives rise to valley-contrasting transport of EM waves~\cite{dongjw2017valley, valley_exp1, valley_exp2}.

The pseudospin and valley degrees of freedom in photonic systems allow the pseudospin- and valley-contrasting transport of EM waves, which has the advantage of unidirectional propagation and easy manipulation by choosing the chirality of the excitation source. Thus it would be highly desirable to introduce both pseudospin and valley degrees of freedom in one photonic system, which would greatly extend the flexibility in manipulating EM energy transport and may provide new opportunities for designing functional EM devices.

\begin{figure*}[htbp]
	\centering
	\includegraphics[width=2\columnwidth]{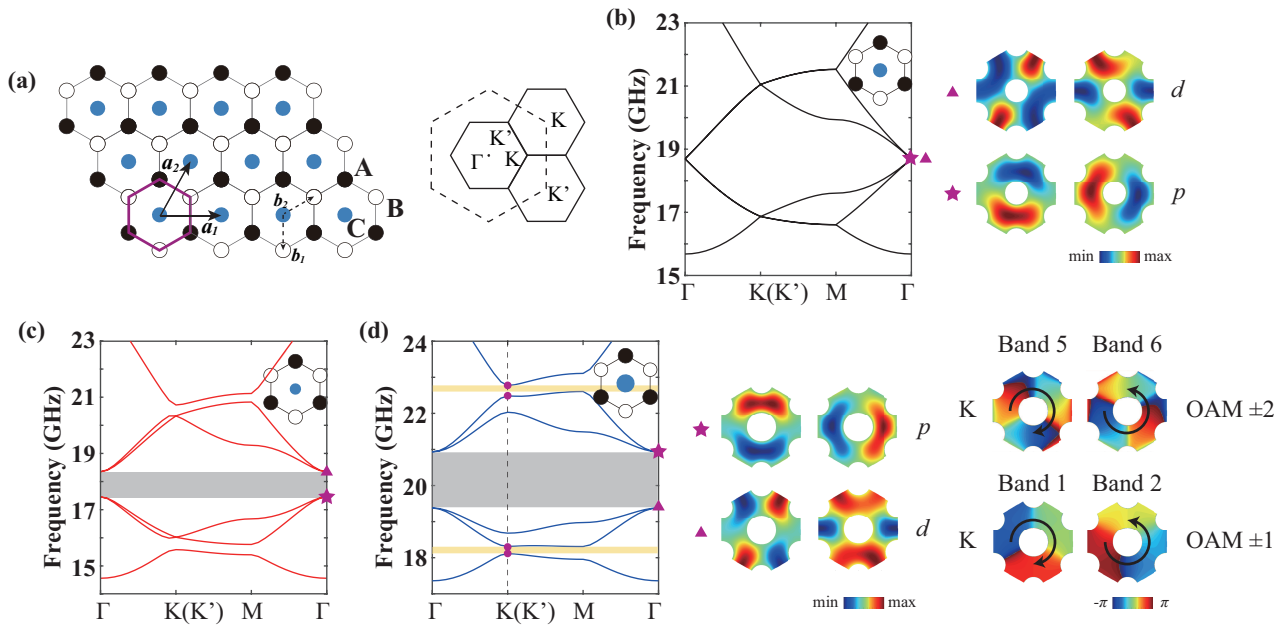}
	\caption{(a) Geometry of the 2D PC arranged in a hybrid triangular lattice with translation vectors of $\bm{a_1}$ and $\bm{a_2}$ and its corresponding BZ (black solid hexagon). Metallic rods with different radii of $r_A, r_B$ and $r_C$ are placed at the lattice sites and the background is a dielectric with dielectric constant of $2.2$.  (b) The band structure and real-space distributions of the electric fields ($E_z$) at the $\Gamma$ point of the band gap edges when $r_A=r_B=r_C=2.5$~mm. (c) The band structure when $r_A=r_B=2.5$~mm, $r_C=2$~mm. (d) The band structure and real-space distributions of $E_z$ at the $\Gamma$ point of the second band gap edges as well as the phase of $E_z$ at the $K$ point of the first and third band gap edges when $r_A=2.6$~mm, $r_B=2.4$~mm, and $r_C=3.2$~mm. Other parameter: lattice constant $a=10\sqrt3$~mm.}
	\label{BG_bulk}
\end{figure*}

In this work, we present a PC structure supporting both the pseudospin and valley degrees of freedom simultaneously. The PC has a hybrid triangular lattice structure, which can be considered as comprising three interlacing triangular sublattices. First, the pseudospin degree of freedom is introduced by increasing the size of one of the three sublattices such that the fourfold degeneracy at the $\Gamma$ point of the BZ is broken, which leads to a topological nontrivial band gap. Next, the sizes of the remaining two sublattices are modified such that the rotation symmetry of the structure is reduced from $C_{6}$ to $C_{3}$ and consequently, the twofold degeneracies at the $K, K'$ points are broken, giving rise to the valley degree of freedom. We note that while the mechanism used in this work to form the pseudospin or valley degree of freedom is known in the literature, combining these two in a single PC is a nontrivial task and has not been studied to the best of our knowledge. Pseudospin- and valley-momentum locking edge states are observed at properly designed domain-wall interfaces and moreover, a four-channel system is presented to demonstrate the novel routing of EM waves that can be selectively controlled by either the pseudospin or valley degree of freedom.

\textit{PC Structure and the Emergence of Nontrivial Pseudospin and Valley Band Gaps.---} We consider a 2D hybrid triangular lattice as shown in Fig.~\ref{BG_bulk}(a), which is formed by three interlacing triangular sublattices A, B and C that share the same lattice vectors $\bm{a_1}$, $\bm{a_2}$ and lattice constant $a$. In general, this hybrid lattice has three-fold rotation symmetry rather than the six-fold rotation symmetry of a regular triangular lattice. For illustration purposes, we utilize metallic rods embedding in a dielectric background to construct the PC. We only consider the TM modes of the system and all the results in this work are obtained by using COMSOL software. Note that due to the boundary conditions of electric conductors, the out-of-plane electric field is only confined in the dielectric region. The idea presented in this work can also be carried over to other PC designs.

When $r_A=r_B=r_C$, there exists a double Dirac cone with four-fold degeneracy at the $\Gamma$ point (Fig.~\ref{BG_bulk}(b)), which can be understood from the band folding phenomenon since the primitive translation vectors now become $\bm{b_1}$ and $\bm{b_2}$ and the $K$ and $K'$ points of the BZ associated with $\bm{b_1}$ and $\bm{b_2}$ will be folded to the $\Gamma$ point of the BZ associated with $\bm{a_1}$ and $\bm{a_2}$. From the eigenstates at the degenerate point shown in the right, we can see there exist two dipole modes and two quadrupole modes. A complete band gap around the double Dirac cone could be induced by simply increasing or decreasing the radius of rods $C$. For example, if we keep the radii of rods at the corners of the primitive hexagonal unit cell (purple hexagon) equal ($r_A=r_B$) and make the size of rods $C$ smaller, a complete band gap opens up around $18$~GHz, as depicted in Fig.~\ref{BG_bulk}(c). After checking the eigenstates at the band gap edges of the $\Gamma$ point, we find two degenerate dipole modes appear at lower eigenfrequency whereas two degenerate quadrupole modes appear at higher eigenfrequency. On the other hand, if we increase the size of rods $C$, a complete band gap appears again as shown in Fig.~\ref{BG_bulk}(d). However, unlike the case in Fig.~\ref{BG_bulk}(c), the locations of the dipole and quadrupole modes are exchanged, indicating a band inversion and thus a topological phase transition.

The PC structure we proposed above is a simple implementation of the hexagonal Su-Schrieffer-Heeger (SSH) model \cite{hex_ssh1,hex_ssh2} or the ``shrink and expand" scheme proposed in \cite{huxiao2015scheme}. As the EM wave is only confined in the dielectric region due to the boundary conditions of electric conductors, six effective ``electromagnetic particles" are formed inside each hexagonal unit cell, see the real-space distributions of the fields shown in Fig.~\ref{BG_bulk}(b) and (d), whose effective hopping between or inside the unit cells can be controlled by the relative size of the radii of rods $A,B$ and $C$. For example, when $r_C<r_A=r_B$, the effective hopping channels inside the unit cell will be increased. As such the intra-hopping will be increased compared to the inter-coupling so that a topological trivial gap is induced, see Fig.~\ref{BG_bulk}(c). On the other hand, when $r_C>r_A \approx r_B$, the inter-hopping becomes stronger than the intra-hopping, thus creating a band inversion between the dipole and quadrupole modes as shown in Fig.~\ref{BG_bulk}(d).

Apart from the band inversion that introduces the pseudospin degree of freedom, valley degree of freedom can also be introduced by making the sizes of rods $A$ and $B$ slightly different so that the Dirac cones at $K$ and $K'$ points (exist at $ r_C>r_A =r_B$, not shown) are gapped out and become valleys, see Fig.~\ref{BG_bulk}(d). The Dirac points at $ r_C>r_A =r_B$ are protected by inversion symmetry and by setting $r_A \neq r_B$, the inversion symmetry is broken and consequently the six-fold rotation symmetry $C_6$ is reduced to $C_3$.  One can see from Fig.~\ref{BG_bulk}(d) that two complete band gaps from $18.1$~GHz to $18.3$~GHz and $22.6$~GHz to $22.8$~GHz appear and the phases of the electric fields ($E_z$) at the $K$ valley band gap edges show intrinsic chirality. In particular, the electric fields carry an angular momentum of order $\pm1$ at the first valley gap edge and $\pm2$ at the second valley gap edge, where the signs are determined by the phase winding directions which are opposite for the upper and lower edges at the same valley. Note that though the valleys at the $K$ and $K'$ points have the same eigenfrequency, they are inequivalent, i.e., compared with the $K$ valley, the phase winding directions at the $K'$ valley edges are reversed (not shown). The emergence of valley states with angular momentum of order $\pm2$ in our PC structure is a notable feature not present in previous works~\cite{dongjw2017valley, multiband}.

\begin{figure}[htbp]
	\centering
	\includegraphics[width=\columnwidth]{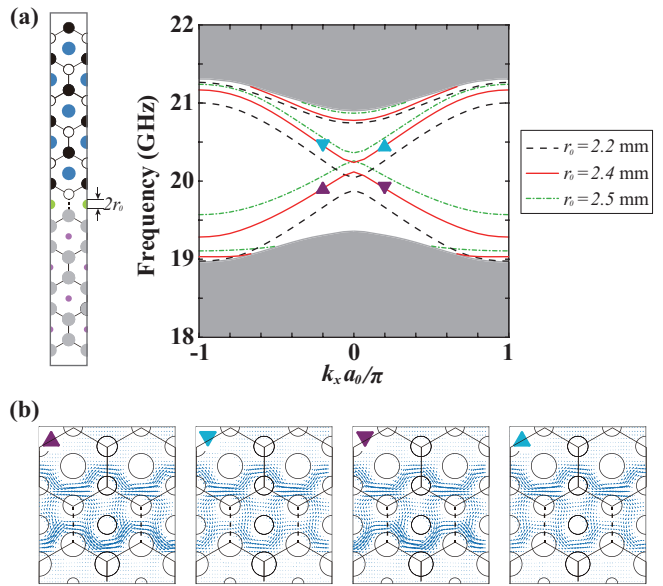}
	\caption{Pseudospin-polarized edge states. (a) Geometry and band structures of the supercell containing a domain wall interface between the topological nontrivial PC shown in Fig.~\ref{BG_bulk}(d) and a trivial PC with parameters of $r_A=r_B=3.1$~mm and $r_C=1.5$~mm, where the parameters of the trivial PC are set to provide the band gap covering of the second band gap of the nontrivial PC. The radius of the green rods at the trivial-nontrivial PC interface is denoted by $r_0$. (b) The time-averaged Poynting vectors at the marked points of the band structure in (a).}
	\label{spin_edge}
\end{figure}

\textit{Pseudospin-Polarized Edge States.---}
Topological edge states can be formed at the interface between two PCs with different topologies according to the bulk-edge correspondence principle. In the following, we will demonstrate the existence of pseudospin-polarized edge states within the second band gap at the interface between the topological nontrivial PC shown in Fig.~\ref{BG_bulk}(d) and a trivial PC with $r_A=r_B>r_C$.

The supercell of the proposed structure is depicted in the left panel of Fig.~\ref{spin_edge}(a). As can be seen from the simulated band structures, there are two bands crossing almost the whole band gap, which are the typical pseudospin-polarized edge states \cite{huxiao2015scheme,menglin_2019TAP}. A novelty of the supercell structure proposed here is that in addition to the rods composing the trivial and nontrivial PCs, there are rods located at the interface (marked by green), which can be used as a useful knob to tune the dispersion of the edge states. For example, from the band structures shown in Fig.~\ref{spin_edge}(a), one can see the dispersion will be shifted toward high frequencies when the radius of these rods $r_0$ increases and vice versa.

\begin{figure}[htbp]
	\centering
	\includegraphics[width=\columnwidth]{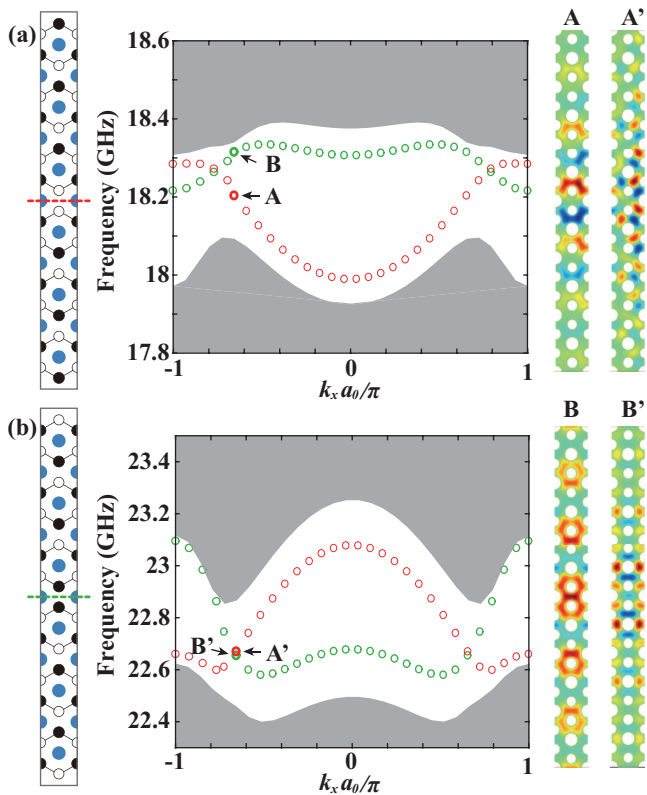}
	\caption{Valley-momentum locking edge states at the domain walls formed with (a) smaller rods and (b) larger rods, and their band structures. The edge states are marked by red for the domain wall in (a) and green for the domain wall in (b). The right panels show the real-space distributions of the electric fields at the marked points in the band structures.}
	\label{valley_edge}
\end{figure}

To further demonstrate the pseudospin-momentum locking behavior of the edge states, we show in Fig.~\ref{spin_edge}(b) the distributions of the time-averaged Poynting vectors at four marked locations of the edge states (see Fig.~\ref{spin_edge}(a)). For all the four states, the EM energy concentrates at the interface between the trivial and nontrivial PCs and flows from one supercell to its adjacent supercell, forming an effective wave-guiding channel. For the states marked by triangles, the net energy flows are along right and for the states marked by the inverted triangles, they are along left. The left- and right-moving paths are accompanied by half-cycle orbits. The rotation direction of the Poynting vectors along the half-cycle orbits correlates with the direction of the net energy flow, implying the pseudospin-momentum locking behavior. We note that this feature is similar to that of the helical edge states found in quantum spin Hall effects \cite{kane_mele1, kane_mele2}.

\textit{Valley-Dependent Edge States.---}
Apart from the second nontrivial band gap as shown in Fig.~\ref{BG_bulk}(d), there exist additionally two band gaps above and below. As there two band gaps are induced by gapping out the corresponding Dirac cones, they are also nontrivial in the sense that the integral of the Berry curvature associated with the frequency band around the two valleys of $K$ and $K'$ gives the valley Chern number of $C_{K,K'} =\pm 1/2 $ \cite{dixiao2007valley, Ma2016all}. Furthermore, the two valleys can be transformed to each other upon the rotation of the PC by $\pi/3, \pi, 5\pi/3$. As a result, valley-momentum locking edge states are supported at the domain wall interface between the PC shown in Fig.~\ref{BG_bulk}(d) and its inversion-symmetric counterpart, i.e., a rotation of $\pi$ with respect to each other, such that the difference of the valley Chern number across the domain-wall interface at $K$ and $K'$ is $+1$ or $-1$.  As the sign of the difference of the valley Chern numbers determines the propagation direction of the emerging edge states, this indicates that the edge state at one valley has a positive velocity whereas the other has a negative one, i.e., the  valley-momentum locking behavior.

To confirm this, we study the band structures of domain walls constructed by the PC of Fig.~\ref{BG_bulk}(d) and its inversion-symmetric counterpart. Note two types of domain wall interfaces can be formed with either the smaller rods or the larger rods at the interface, see Fig.~\ref{valley_edge} (a) and (b). For each domain wall, there is one edge state within each valley band gap above and below. The group velocities of the edge states at different valleys are opposite, indicating the valley-momentum locking behavior. Furthermore, from the real-space distributions of the electric fields ($E_z$), we can see that the edge states show different symmetries at the two domain wall interfaces: they are asymmetric for the interface with smaller rod and symmetric for the interface with larger rods within both the valley band gaps above and below.

\begin{figure}[htbp]
	\centering
	\includegraphics[width=\columnwidth]{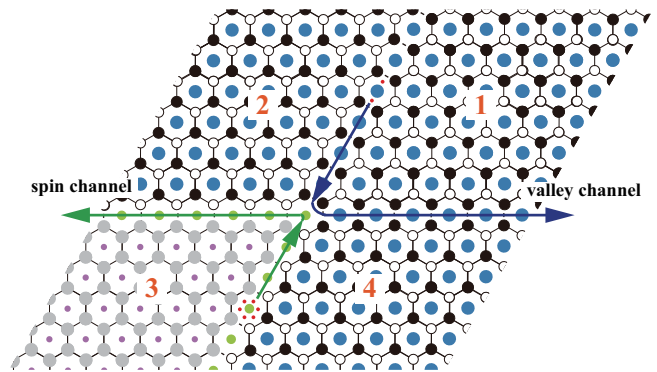}
	\caption{A four-channel system for EM routing that consists of a $60$-degree trivial-nontrivial PC interface as in Fig.~\ref{spin_edge}(a) and another $60$-degree interface in Fig.~\ref{valley_edge}(b). The six-line-source array and two-line-source array for the excitation in the numerical model are also shown.}
	\label{4ch}
\end{figure}

\textit{Excitation of the Pseudopin- and Valley-Hall like Edge States.---}
The pseudospin- and valley-momentum locking edge states discussed above could be useful for designing functional EM devices. To demonstrate this, we build a four-channel system (see Fig.~\ref{4ch}) that consists of two EM routing channels, which would be selectively activated by either the pseudospin or valley degree of freedom. As the propagation direction of the pseudospin-polarized edge states is determined by its inherent chirality, i.e., the rotation direction of the energy flow shown in Fig.~\ref{spin_edge}(b), the two unidirectional propagations could be selectively excited by a chiral source with positive or negative angular momentum. We utilize a six-line-source array that generates an angular momentum of order $1$ within the hexagonal cluster at the interface for excitation in COMSOL. The line sources are located equidistantly on the perimeter of a circle. They are assigned with the same weight but different phases, ranging from zero to $5\pi/3$ with a successive anticlockwise increment of $\pi/3$. On the other hand, the valley-dependent edge states show distinct eigenfield distributions at different valleys. Therefore, two line sources with the same magnitude and designed phase difference that coincides with the eigenstates are used for the excitation of the state at the selected valley within both band gaps.

\begin{figure}[htbp]
	\centering
	\includegraphics[width=\columnwidth]{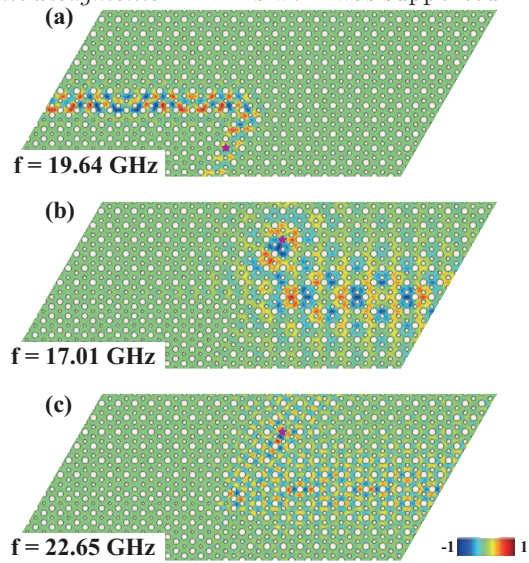}
	\caption{The simulated distributions of the edge states. (a) The excited pseudospin-polarized edge state by a six-line-source array carrying OAM of order $1$ at $19.64$~GHz. The excited valley-dependent edge state at (b) $17.01$~GHz (in the first band gap) and (c) $22.65$~GHz (in the third band gap). The nontrivial PC has the parameters of $r_A=2.2$~mm, $r_B=2$~mm, $r_C=3.5$~mm. The trivial PC has the parameters of $r_A=r_B=3.1$~mm and $r_C=1.5$~mm, and the rods at the trivial-nontrivial PC interface have the radius of $2.5$~mm.}
	\label{simE}
\end{figure}

The simulated real-space distributions of $E_z$ within the three band gaps are depicted in Fig.~\ref{simE}. The parameters for the PCs here are changed to enlarge the valley band gaps, which will not alter the physics we proposed. At $19.64$~GHz, the pseudospin-polarized edge state is successfully excited by the chiral source. The field is well confined at the trivial-nontrivial PC interface and propagates only in the direction correlating with the chirality of the source without being backscattered by the sharp bend. Meanwhile, we can also observe the unidirectional propagation for the edge states within the valley band gaps. The field distributions within the two band gaps before and after the sharp bend are consistent with those in Fig.~\ref{valley_edge}(b). It is worth noting that while the unidirectional propagation of the pseudospin-polarized edge states at other frequencies can be excited using the same settings of the source, the valley-dependent edge states can only be unidirectionally excited by adjusting the two line sources accordingly: the phase difference between the two line sources should be set in accordance with the eigenstate at the selected frequency. In addition to the capability of maintaining unidirectional propagation against sharp bends, the pseudospin- and valley-Hall like edge states are also robust against moderate disorders. The robustness of individual effects against lattice and structure disorders has been analyzed in~\cite{huxiao2015scheme,valley_circuit2,dong2020tunable}, which implies similar behavior of our proposed structure.

For experimental realization, the 2D model system can be implemented by 3D structure with the array of metallic rods inserted between two metallic parallel plates, the response of which shows a good agreement with the numerical results~\cite{huxiao2015scheme,qsh_18PRL}. Furthermore, for ease of implementation, the metallic rods can be replaced by metallic vias so that the whole design can be realized by drilling metallic vias into a dielectric material with metal coating on its top and bottom surfaces.

To use only the pseudospin- (valley-) polarized wave-guiding channel, one can set the upright (downleft) quadrant in Fig.~\ref{4ch} identical to those of the upleft and downright quadrants. On the other hand, to exploit both the pseudospin- and valley-Hall like edge states, two wave-guiding channels as demonstrated in Fig.~\ref{4ch} could be built. One could further introduce more channels by dividing the PC structure into multiple domains as studied in \cite{valley_network}. Therefore, a compact tri-band and multichannel wave-guiding system can simply be realized with the pseudospin and valley degrees of freedom, which could potentially be used as optical multiplexing/demultiplexing devices where multiband and multichannel response is required.

\textit{Conclusion and outlook.---}
In conclusion, we have demonstrated the coexistence of pseudospin- and valley-Hall-like edge states in a PC structure with $C_{3v}$ symmetry, which is composed of three interlacing triangular sublattices. In the proposed structure, a band inversion at the $\Gamma$ point induced by simply changing the size of the rods in one sublattice creates a topological nontrivial band gap that encodes the pseudospin physics. Furthermore, changing the sizes of the rods in the remaining two sublattices leads to the gapping out of the Dirac points at the $K, K'$ points which introduces the valley physics. The existence of pseudospin- and valley-momentum locking edge states at properly designed interfaces are demonstrated through dispersion analysis. To demonstrate the usefulness of the system with both the pseudospin and valley degrees of freedom, we have designed a four-channel system, where the unidirectional routing of EM waves against sharp bends between two routes can be selectively excited by either the pseudospin or the valley degree of freedom. Beyond the potential multiplexing/demultiplexing application explained in the main text, this multiband and multi-degree of freedom system could also be potentially useful in a range of topological photonic applciations, such as topological photonic integrated circuits \cite{valley_circuit1, valley_circuit2}, high-performance lasing \cite{valley_lasing1, valley_lasing2, spin_lasing}, chiral light-matter interactions \cite{emitter1, emitter2, emitter3, emitter4} and nonlinear topological photonics \cite{lan_ntp, you_ntp, yuri_ntp}.

\textit{Acknowledgments. ---}
This work was supported in part by the Research Grants Council of Hong Kong GRF 17209918, AOARD FA2386-17-1-0010, NSFC 61271158,  NSFC 61975177, HKU Seed Fund 201711159228, and Thousand Talents Program for Distinguished Young Scholars of China.


\begin{thebibliography}{98}

\bibitem{topo_wen} X.-G. Wen, {\it Choreographed entanglement dances: Topological states of quantum matter}, Science {\bf 363}, eaal3099 (2019).
\bibitem{QH_klitzing17} K. v. Klitzing, {\it Quantum Hall Effect: Discovery and Application},  Annu. Rev. Condens. Matter Phys. {\bf 8}, 13 (2017).

\bibitem{kane2010colloquium} M. Z. Hasan and C. L. Kane, {\it Colloquium: Topological insulators}, Rev. Mod. Phys. {\bf 82}, 3045 (2010).
\bibitem{sczhang2011TI} X.-L. Qi and S.-C. Zhang, {\it Topological insulators and superconductors}, Rev. Mod. Phys. {\bf 83}, 1057 (2011).
\bibitem{review_RMP19}T. Ozawa, H. M. Price, A. Amo, N. Goldman, M. Hafezi, L. Lu, M. C. Rechtsman, D. Schuster, J. Simon, O. Zilberberg, and I. Carusotto, {\it Topological photonics}, Rev. Mod. Phys. {\bf 91}, 015006 (2019).

\bibitem{qh80}K. v. Klitzing, G. Dorda, and M. Pepper, {\it New Method for High-Accuracy Determination of the Fine-Structure Constant Based on Quantized Hall Resistance}, Phys. Rev. Lett. {\bf 45}, 494 (1980).
\bibitem{kane_mele1}C. L. Kane and E. J. Mele, {\it Quantum Spin Hall Effect in Graphene}, Phys. Rev. Lett. {\bf 95}, 226801 (2005).
\bibitem{kane_mele2} C. L. Kane and E. J. Mele, {\it $Z_2$ Topological Order and the Quantum Spin Hall Effect}, Phys. Rev. Lett. {\bf 95}, 146802 (2005).
\bibitem{dixiao2007valley} D. Xiao, W. Yao, and Q. Niu, {\it Valley-Contrasting Physics in Graphene: Magnetic Moment and Topological Transport}, Phys. Rev. Lett. {\bf 99}, 236809 (2007).

\bibitem{joannopoulos1997PCs} J. D. Joannopoulos, P. R. Villeneuve, and S. Fan, {\it Photonic crystals: putting a new twist on light},  Nature {\bf 386}, 143 (1997).

\bibitem{haldane08} F. D. M. Haldane and S. Raghu, {\it Possible Realization of Directional Optical Waveguides in Photonic Crystals with Broken Time-Reversal Symmetry}, Phys. Rev. Lett. {\bf 100}, 013904 (2008).
\bibitem{Soljacic08}Z. Wang, Y. D. Chong, J. D. Joannopoulos, and M. Soljacic, {\it Reflection-Free One-Way Edge Modes in a Gyromagnetic Photonic Crystal}, Phys. Rev. Lett. {\bf 100}, 013905 (2008).
\bibitem{Soljacic09} Z. Wang, Y. Chong, J. D. Joannopoulos, and M. Soljacic, {\it Observation of unidirectional backscattering-immune topological electromagnetic states}, Nature {\bf 461}, 772 (2009).

\bibitem{Khanikaev2013photonic} A. B. Khanikaev, S. H. Mousavi, W.-K. Tse, M. Kargarian, A. H. MacDonald, and G. Shvets, {\it Photonic topological insulators}, Nat. Mater. {\bf 12}, 233 (2013).
\bibitem{chan_nc14} W.-J. Chen, S.-J. Jiang, X.-D. Chen, B. Zhu, L. Zhou, J.-W. Dong, and C. T. Chan, {\it Experimental realization of photonic topological insulator in a uniaxial metacrystal waveguide}, Nat. Commun. {\bf 5}, 5782 (2014).
\bibitem{Ma2015guiding} T. Ma, A. B. Khanikaev, S. H. Mousavi, and G. Shvets, {\it Guiding electromagnetic waves around sharp corners: Topologically protected photonic transport in metawaveguides}, Phys. Rev. Lett. {\bf 114}, 127401 (2015).
\bibitem{chen_pnas} C. He, X.-C. Sun, X.-P. Liu, M.-H. Lu, Y. Chen, L. Feng, and Y.-F. Chen, {\it Photonic topological insulator with broken time-reversal symmetry}, Proc. Natl. Acad. Sci. U.S.A {\bf 113}, 4924 (2016).
\bibitem{Khanikaev2016robust} X. Cheng, C. Jouvaud, X. Ni, S. H. Mousavi, A. Z. Genack, and A. B. Khanikaev, {Robust reconfigurable electromagnetic pathways within a photonic topological insulator}, Nat. Mater. {\bf 15}, 542 (2016).
\bibitem{tetm_apl19} A. Slobozhanyuk,   A. V. Shchelokova, X. Ni, S. H. Mousavi, D. A. Smirnova, P. A. Belov, A. Alu, Y. S. Kivshar, and A. B. Khanikaev, {\it Near-field imaging of spin-locked edge states in all-dielectric topological metasurfaces}, Appl. Phys. Lett. {\bf 114}, 031103 (2019).
\bibitem{Sievenpiper_lpr19} D. J. Bisharat and D. F. Sievenpiper, {\it Electromagnetic-Dual Metasurfaces for Topological States along a 1D Interface}, Laser Photonics Rev. {\bf 13}, 1900126 (2019).

\bibitem{huxiao2015scheme} L.-H. Wu and X. Hu, {\it Scheme for achieving a topological photonic crystal by using dielectric material}, Phys. Rev. Lett. {\bf 114}, 223901 (2015).
\bibitem{qsh_18PRL}Y. Yang, Y. F. Xu, T. Xu, H.-X. Wang, J.-H. Jiang, X. Hu, and Z. H. Hang, {\it Visualization of a Unidirectional Electromagnetic Waveguide Using Topological Photonic Crystals Made of Dielectric Materials}, Phys. Rev. Lett. {\bf 120}, 217401 (2018).
\bibitem{qsh_19PRL}S. Peng, N. J. Schilder, X. Ni, J. van de Groep, M. L. Brongersma, A. Alu, A. B. Khanikaev, H. A. Atwater, and A. Polman, {\it Probing the Band Structure of Topological Silicon Photonic Lattices in the Visible Spectrum}, Phys. Rev. Lett. {\bf 122}, 117401 (2019).
\bibitem{qsh_20sciadv} N. Parappurath, F. Alpeggiani, L. Kuipers, and E. Verhagen, {\it Direct observation of topological edge states in silicon photonic crystals: Spin, dispersion, and chiral routing}, Sci. Adv. {\bf 6}, eaaw4137 (2020).
\bibitem{qsh_20nanoletter} W. Liu, M. Hwang, Z. Ji, Y. Wang, G. Modi, and R. Agarwal, {\it $Z_2$ Photonic Topological Insulators in the Visible Wavelength Range for Robust Nanoscale Photonics}, Nano Lett. {\bf 20}, 1329 (2020).


\bibitem{Ma2016all} T. Ma and G. Shvets, {\it All-Si valley-Hall photonic topological insulator}, New J. Phys. {\bf 18}, 025012 (2016).
\bibitem{dongjw2017valley}  X.-D. Chen, F.-L. Zhao, M. Chen, and J.-W. Dong, {\it Valley-contrasting physics in all-dielectric photonic crystals: Orbital angular momentum and topological propagation}, Phys. Rev. B {\bf 96}, 020202 (2017).

\bibitem{valley_exp1}X.-T. He, E.-T. Liang, J.-J. Yuan, H.-Y. Qiu, X.-D. Chen, F.-L. Zhao, and J.-W. Dong, {\it A silicon-on-insulator slab for topological valley transport}, Nat. Commun. {\bf 10}, 872 (2019).
\bibitem{valley_exp2} M. I. Shalaev, W. Walasik, A. Tsukernik, Y. Xu, and N. M. Litchinitser, {\it Robust topologically protected transport in photonic crystals at telecommunication wavelengths}, Nat. Nanotechnol. {\bf 14}, 31 (2019).


\bibitem{hex_ssh1}L.-H. Wu and X. Hu, {\it Topological Properties of Electrons in Honeycomb Lattice with Detuned Hopping Energy}, Sci. Rep. {\bf 6}, 24347 (2016).
\bibitem{hex_ssh2} F. Liu, H.-Y. Deng, and K. Wakabayashi, {\it Helical Topological Edge States in a Quadrupole Phase}, Phys. Rev. Lett. {\bf 122}, 086804 (2019).

\bibitem{multiband} Q. Chen, L. Zhang, M. He, Z. Wang, X. Lin, F. Gao, Y. Yang, B. Zhang, and H. Chen, {\it Valley-Hall Photonic Topological Insulators with Dual-Band Kink States}, Adv. Optical Mater. {\bf 7}, 1900036 (2019).

\bibitem{menglin_2019TAP} M. L. N. Chen, L. J. Jiang, Z. Lan, and W. E. I. Sha. {\it Pseudospin-Polarized Topological Line Defects in Dielectric Photonic Crystals}, IEEE Trans. Antennas Propag. {\bf 68}, 609 (2020).


\bibitem{dong2020tunable} Z. Dong, F. Xu, and W. Liang,  {\it A tunable silicon-on-insulator valley Hall photonic crystal at telecommunication wavelengths}, EPL (Europhysics Letters) {\bf 131}, 54002 (2020).




\bibitem{valley_circuit2} L. Zhang, Y. Yang, M. He, H.-X. Wang, Z. Yang, E. Li, F. Gao, B. Zhang, R. Singh, J.-H. Jiang, and H. Chen, {\it Valley Kink States and Topological Channel Intersections in Substrate-Integrated Photonic Circuitry}, Laser Photonics Rev. {\bf 13}, 1900159 (2019).

\bibitem{valley_network} M. P. Makwana and R. V. Craster, {\it Designing multidirectional energy splitters and topological valley supernetworks}, Phys. Rev. B {\bf 98}, 235125 (2018).

\bibitem{valley_circuit1} J. Ma, X. Xi, and X. Sun, {\it Topological Photonic Integrated Circuits Based on Valley Kink States}, Laser Photonics Rev. {\bf 13}, 1900087 (2019).



\bibitem{valley_lasing1} Y. Zeng, U. Chattopadhyay, B. Zhu, B. Qiang, J. Li, Y. Jin, L. Li, A. G. Davies, E. H. Linfield, B. Zhang, Y. Chong, and Q. J. Wang, {\it Electrically pumped topological laser with valley edge modes},  Nature {\bf 578}, 246 (2020).
\bibitem{valley_lasing2} Y. Gong, S. Wong, A. J. Bennett, D. L. Huffaker, and S. S. Oh, {\it Topological Insulator Laser Using Valley-Hall Photonic Crystals}, ACS Photonics {\bf 7}, 2089 (2020).
\bibitem{spin_lasing} Z.-K. Shao, H.-Z. Chen, S. Wang, X.-R. Mao, Z.-Q. Yang, S.-L. Wang, X.-X. Wang, X. Hu, and R.-M. Ma, {\it A high-performance topological bulk laser based on band-inversion-induced reflection}, Nat. Nanotechnol. {\bf 15}, 67 (2020).



\bibitem{emitter1}S. Barik, A. Karasahin, C. Flower, T. Cai, H. Miyake, W.  DeGottardi, M. Hafezi, and E. Waks, {\it A topological quantum optics interface}, Science {\bf 359}, 666 (2018).
\bibitem{emitter2} T. Yamaguchi, Y. Ota, R. Katsumi, K. Watanabe, S. Ishida, A. Osada, Y. Arakawa, and S. Iwamoto, {\it GaAs valley photonic crystal waveguide with light-emitting InAs quantum dots}, Appl. Phys. Express {\bf 12}, 062005 (2019).
\bibitem{emitter3} M. J. Mehrabad, A. P. Foster, R. Dost, A. M. Fox, M. S. Skolnick, and L. R. Wilson, {\it Chiral topological photonics with an embedded quantum emitter}, arXiv:1912.09943.
\bibitem{emitter4} S. Barik, A. Karasahin, S. Mittal, E. Waks, and M. Hafezi, {\it Chiral quantum optics using a topological resonator}, Phys. Rev. B {\bf 101}, 205303 (2020). 

\bibitem{lan_ntp} Z. Lan, J. W. You, and N. C. Panoiu, {\it Nonlinear one-way edge-mode interactions for frequency mixing in topological photonic crystals}, Phys. Rev. B {\bf 101}, 155422 (2020).
\bibitem{you_ntp} J. W. You, Z. Lan, and N. C. Panoiu, {\it Four-wave mixing of topological edge plasmons in graphene metasurfaces}, Sci. Adv. {\bf 6}, eaaz3910 (2020).
\bibitem{yuri_ntp}D. Smirnova, D. Leykam, Y. Chong, and Y. Kivshar, {\it Nonlinear topological photonics}, Appl. Phys. Rev. {\bf 7}, 021306 (2020).




\end{thebibliography}
\end{document}